\documentclass[a4paper,twoside]{article}

\usepackage{epsfig}
\usepackage{subfig}
\usepackage{calc}
\usepackage{amssymb}
\usepackage{amstext}
\usepackage{amsmath}
\usepackage{amsthm}
\usepackage{multicol}
\usepackage{pslatex}
\usepackage{apalike}
\usepackage{graphicx}
\usepackage{amsfonts}
\usepackage{booktabs}
\usepackage{algorithm}
\usepackage{algorithmic}
\usepackage{xcolor}
\usepackage{url}

\usepackage{SCITEPRESS}     

\newcommand{\cL}{\mathcal{L}}
\newcommand{\vc}[1]{\boldsymbol{#1}}

\begin{document}

\title{Evolutionary techniques in lattice sieving algorithms}

\author{\authorname{Thijs Laarhoven \sup{1}\orcidAuthor{0000-0002-2369-9067}}
\affiliation{\sup{1}Eindhoven University of Technology}
\email{mail@thijs.com}
}


\abstract{Lattice-based cryptography has recently emerged as a prominent candidate for secure communication in the quantum age. Its security relies on the hardness of certain lattice problems, and the inability of known lattice algorithms, such as lattice sieving, to solve these problems efficiently. In this paper we investigate the similarities between lattice sieving and evolutionary algorithms, how various improvements to lattice sieving can be viewed as applications of known techniques from evolutionary computation, and how other evolutionary techniques can benefit lattice sieving in practice.}

\onecolumn \maketitle \normalsize \setcounter{footnote}{0} \vfill

\section{INTRODUCTION}
\label{sec:intro}

\paragraph{Cryptography.} To protect digital communication between two parties against eavesdroppers, techniques from the field of \textit{cryptography} are widely used, ensuring that only the legitimate parties are able to extract the contents of the exchanged messages. Although most of the currently deployed systems, such as RSA encryption~\cite{rivest78} and Diffie--Hellman key exchange~\cite{diffie76}, are considered reasonably efficient and secure against ``classical'' adversaries, a breakthrough work of Shor~\cite{shor97} demonstrated that most existing solutions are completely insecure when quantum computers become a reality -- even if building an efficient quantum computer may still be decades away, nothing stops adversaries from storing encrypted communication with classical technologies now, and decrypting the contents when a quantum computer has been built. Classical cryptographic methods therefore pose a risk even today, and academia and industry worldwide are increasingly shifting their attention towards new, quantum-proof cryptographic primitives~\cite{etsi19,nist17}.

\paragraph{Lattice-based cryptography and cryptanalysis.} Among the proposed methods for quantum-safe cryptography, \textit{lattice-based cryptography} has established itself as a leading candidate, offering versatile, advanced, and efficient cryptographic designs with small key sizes~\cite{regev05,micciancio09}. Its security relies on the hardness of certain lattice problems, such as the shortest and closest vector problems, and a crucial aspect of designing lattice-based cryptographic primitives is \textit{cryptanalysis}: analyzing the security of these schemes, and accurately assessing the hardness of the underlying problems. After all, overestimating the true costs of solving these problems would lead to overly optimistic security estimates and insecure schemes, while underestimating these costs would lead to unnecessarily large parameters. In practice, the only way to accurately choose parameters and to assess the hardness of these problems is to consider state-of-the-art algorithms for solving these problems, and estimating their costs for large parameters. 

\paragraph{Lattice sieving.} Currently, the fastest known method for solving most hard lattice problems is \textit{lattice sieving}~\cite{ajtai01}. This method has the best known scaling of the time complexity with the lattice dimension. Asymptotic cost estimates for sieving of~\cite{becker16lsf,laarhoven16phd} have now been extensively used for choosing parameters in various lattice-based cryptographic schemes; see e.g.~\cite{alkim16,crystals2}. Until recently lattice sieving was often not considered as practical for low dimensions as \textit{lattice enumeration}~\cite{gama10}, but recent work has truly demonstrated the superiority of lattice sieving in practice as well~\cite{albrecht19,svp}. Given the above, it is crucial that we obtain a good understanding of lattice sieving algorithms, and how they fit in the bigger picture of algorithms in general.

\subsection{Contributions}

In this paper we describe how lattice sieving can naturally be viewed as an evolutionary algorithm, and we describe how various improvements to lattice sieving from recent years can be traced back to closely related computational techniques in dealing with evolving populations. Besides describing this novel connection, and investigating this new relation between AI and cryptography, we list opportunities for lattice sieving to benefit from techniques known from the field of evolutionary algorithms, and experimentally assess some of these for their impact on the performance on state-of-the-art lattice sieving methods\footnote{A previous paper~\cite{ding15} attempted to use genetic techniques for solving hard lattice problems, but attempted to apply these techniques to lattice enumeration, and did not succeed in obtaining an efficient, competitive algorithm. The more natural and novel relation with lattice sieving is explored here.}. We further briefly discuss the main results, and provide ideas for future research on this intersection of artificial intelligence and cryptography.


\paragraph{Outline.} The remainder of the paper is organized as follows. In Section~\ref{sec:pre} we briefly cover notation and some preliminaries on lattices and lattice sieving algorithms. Section~\ref{sec:evol} describes (basic) lattice sieving in the framework of evolutionary algorithms. Section~\ref{sec:tech1} covers previous advances in lattice sieving, and how they relate to techniques from evolutionary computation, while Section~\ref{sec:tech2} covers those techniques from the field of AI that have not yet been applied to lattice sieving. Section~\ref{sec:exp} describes experiments with these techniques, and Section~\ref{sec:disc} concludes with a discussion on the relation between these fields, and avenues for further exploration.

\section{PRELIMINARIES}
\label{sec:pre}

\paragraph{Lattices.} Mathematically speaking, a $d$-dimensional lattice represents a discrete subgroup of $\mathbb{R}^d$. More concretely, given a set of $d$ independent vectors $\vc{B} = \{\vc{b}_1, \dots, \vc{b}_d\} \subset \mathbb{R}^d$, we define the lattice generated by the basis $\vc{B}$ as follows:
\begin{align}
    \cL = \cL(\vc{B}) := \left\{\sum_{i=1}^d \lambda_i \vc{b}_i : \lambda_i \in \mathbb{Z}\right\}.
\end{align}
Intuitively, working with lattices can be seen as doing linear algebra over the integers; a point is in the lattice if and only if it can be described by an \textit{integer} linear combination of the basis vectors. A crucial property of lattices for algorithms considered in this paper is that if both $\vc{v}, \vc{w} \in \cL$, then also $m \vc{v} + n \vc{w} \in \cL$ for $m, n \in \mathbb{Z}$: we can combine lattice vectors to form new lattice vectors. Examples of lattices include $\mathbb{Z}^d$ (all integer vectors), and the 2D hexagonal lattice.

\paragraph{The shortest vector problem (SVP).} Although describing how lattice-based cryptography works falls outside the scope of this paper (interested readers may refer to e.g.~\cite{regev06,micciancio09}), an important aspect of this area of cryptography is that its security relies on the hardness of lattice problems such as the \textit{shortest vector problem (SVP)}. Given a description of a lattice, this problem asks to find a non-zero vector $\vc{s} \in \cL$ with smallest Euclidean norm: \begin{align}
    \|\vc{s}\| = \min_{\vc{0} \neq \vc{v} \in \cL} \|\vc{v}\|. \qquad \left(\|\vc{x}\|^2 := \textstyle\sum_{i=1}^d x_i^2\right)
\end{align} 
Although for e.g.\ $\cL = \mathbb{Z}^d$ this problem is easy, for random lattices this problem is known to be hard~\cite{khot04b}, and becomes increasingly difficult as the dimension $d$ increases. Currently the fastest known methods for solving SVP in high dimensions are based on \textit{lattice sieving}, and asymptotically the fastest algorithm has a time complexity scaling as $2^{0.292d + o(d)}$ classically~\cite{becker16lsf}, or $2^{0.265d + o(d)}$ when using quantum computers~\cite{laarhoven16phd}.\footnote{Recall that lattice-based cryptography is advertised as quantum-safe, even though faster quantum algorithms exist for solving these problems than with classical computers. This is because the scaling of the best time complexity remains exponential in $d$, whereas for e.g.\ RSA or discrete logarithm settings, the best known attack costs decrease from (sub)exponential to only polynomial in the security parameter when using a quantum computer~\cite{shor97}.} Benchmarks on random lattices~\cite{svp} further demonstrate the practicality of sieving algorithms up to dimensions $d \approx 150$. Beyond these dimensions, the time and memory complexities are too large for academic testing~\cite{albrecht19}.  


\begin{algorithm}[tb]
\caption{Outline of evolutionary algorithms}
\label{alg:evo}
\begin{algorithmic}[1] 
\STATE Initialize a population with random candidate solutions
\STATE Evaluate each candidate for their fitness
\REPEAT
    \STATE Select parents for breeding
    \STATE Recombine parents to form new children
    \STATE Mutate some of the resulting offspring
    \STATE Evaluate new candidates for their fitness
    \STATE Select individuals for the next generation
\UNTIL{A termination condition is satisfied}
\end{algorithmic}
\end{algorithm}

\paragraph{Evolutionary algorithms (EAs).} Let us briefly also recall the basics of evolutionary algorithms. These algorithms model the flow of evolution as found in nature, where a population evolves and adapts to its environmental circumstances to allow for optimal survival rates of the species. Algorithm~\ref{alg:evo} presents pseudo-code of evolutionary algorithms, which model this process from an algorithmic point of view, and the key components are briefly discussed below.

\begin{description}
    \item[Initialization.] Initially a random sample of members from the species is generated. The stronger the initial population, the less time the population needs to evolve to its optimal form, but commonly the initialization process is less important than the regenerational steps.
    \item[Parent selection.] Ideally, individual members of a population should breed only if this process is likely to result in genetically strong offspring. For this, parents should be selected based on some (fitness) criteria that guarantees that this will often be the case.
    \item[Recombination.] Given two members of the population, recombination (crossover) defines a method for (randomly) generating a child from these two parents. This child should inherit fitness properties from its parents, to guarantee that the species improves over time. 
    \item[Mutation.] Mutations of individual members of the population stimulate genetic diversity, preventing an early convergence to local optima. Mutations are not always present in evolutionary algorithms, and if present often only occur with very small probability.
    \item[Fitness evaluation.] In most settings, the population has a certain goal, e.g.\ to survive for as long as possible, or to bring individuals closer to an optimal solution. For this it is important to be able to test the fitness of individual members of the population. Here we will restrict ourselves to \textit{absolute} fitness functions, where individuals can be ranked based on their fitness levels.
    \item[Survivor selection.] In evolutionary algorithms, generally only the fittest members of the population are allowed to survive between generations (survival of the fittest). This is often enforced by selecting the subset of individuals with the highest fitness levels for the next generation, and discarding the remainder of the population.
\end{description}
For more on commonly used techniques in evolutionary algorithms, we refer the interested reader to e.g.\ \cite{back96,back00,back00b,coello07}.


\section{LATTICE SIEVING AS AN EA}
\label{sec:evol}

As outlined above, lattice sieving is currently the fastest method for solving problems such as SVP, and understanding this algorithm well is therefore crucial for accurately designing efficient quantum-secure communication protocols. Understanding its relation with techniques from other fields may prove useful as well, and might inspire further advances in lattice sieving in the future. One of our main contributions is studying this relation between lattice sieving and evolutionary algorithms, and below we will show how lattice sieving can naturally be phrased as an evolutionary method.

\paragraph{Original description.} As a starting point, let us take the Nguyen--Vidick sieve~\cite{nguyen08}, which was the first practical, heuristic lattice sieving algorithm for solving SVP. This method starts by sampling a list $P$ of exponentially many random, rather long lattice vectors, e.g.\ by taking small, random integer linear combinations of the basis vectors. The idea of lattice sieving is then to iteratively apply a \textit{sieve} to $P$ to form a new list $P'$ of shorter lattice vectors, which will then replace $P$. For this, the algorithm considers all pairs of vectors $\vc{v}, \vc{w} \in P$, and sees whether $\vc{u} = \vc{v} - \vc{w}$ forms a shorter lattice vector than either $\vc{v}$ or $\vc{w}$. If this is the case, we keep $\vc{u}$ for the next list $P'$, and we discard the longest of $\vc{v}, \vc{w}$ for the next generation. After repeatedly applying this sieve, the vectors in the list become shorter and shorter, and ultimately the list $P$ will contain most of the shortest vectors in the lattice, including the solution. 

\paragraph{As an evolutionary algorithm.} This algorithm can naturally be viewed as an evolutionary process, and Algorithm~\ref{alg:evosieve} describes this method in pseudo-code. Below we briefly describe how the previously listed key components of evolutionary methods appear in lattice sieving.
\begin{description}
    \item[Initialization.] Commonly, the initial population of lattice vectors is generated by using a discrete Gaussian sampler over the lattice, guaranteeing that (1) these vectors can be sampled quickly, and (2) the resulting sampled vectors are not unnecessarily long.
    \item[Parent selection.] Two members (vectors) $\vc{v}, \vc{w}$ from the population are considered for breeding if they are relatively nearby in space, so that $\vc{u} = \vc{v} - \vc{w}$ is shorter than at least one of its parents.
    \item[Recombination.] Then, given that $\vc{v}, \vc{w}$ are appropriate parents for reproduction, recombination deterministically results in the child $\vc{u} = \vc{v} - \vc{w} \in \cL$, which is also in the lattice and is hopefully a shorter lattice vector.
    \item[Mutation.] Existing lattice sieving methods do not include mutations; offspring is deterministically generated from the parents, and no individual mutations take place.
    \item[Fitness evaluation.] As the goal of the algorithm is to obtain short (non-zero) lattice vectors, a natural measure for the fitness of individual population members $\vc{v} \in P$ is their Euclidean norm $\|\vc{v}\|$: the smaller this norm, the fitter. 
    \item[Survivor selection.] After offspring has been produced, selections are made locally: if a child has a lower norm than one of its parents, we discard the longest of the parent vectors and replace it with the child.
\end{description}


\paragraph{Similarities and differences.} Lattice sieving can naturally be viewed as an evolutionary algorithm, as the core procedure consists of generating a large population, and doing simple, local recombinations on pairs of vectors to form shorter lattice vectors, which replace the longer ones in the initial population. The actual computational operations in sieving rely on the following very elementary property of lattices: 
\begin{align}
    \vc{v}, \vc{w} \in \cL \quad \implies \quad \vc{v} - \vc{w} \in \cL
\end{align}
By only recombining suitable pairs of parents, we guarantee that the members of the population become increasingly fit.

Arguably the biggest differences compared to the standard evolutionary model are that (1) mutations do not exist at all in existing lattice sieving approaches; and (2) parents are directly replaced by their children, rather than the survivor selection happening on a population-wide, global scale. Note also that recombination and offspring generation happen deterministically -- there is no randomness in computing the child $\vc{u} = \vc{v} - \vc{w}$ from the parents $\vc{v}, \vc{w}$.

\begin{algorithm}[tb]
\caption{Evolutionary lattice sieving}
\label{alg:evosieve}
\textbf{Input}: A description (basis) of a lattice $\cL$ \\
\textbf{Output}: A population $P \subset \cL$ of short lattice vectors\\ \vspace{-0.4cm}
\begin{algorithmic}[1] 
\STATE Initialize a population $P \subset \cL$ of random lattice vectors
\REPEAT
    \FORALL{Potential parents $\vc{v}, \vc{w} \in P$}
        \STATE Generate the (potential) offspring $\vc{u} = \vc{v} - \vc{w}$
        \IF{$\|\vc{u}\| < \|\vc{v}\|$ \textbf{and/or} $\|\vc{u}\| < \|\vc{w}\|$}
            \STATE Replace the longest parent with $\vc{u}$
        \ENDIF
    \ENDFOR
\UNTIL{$P$ contains sufficiently short lattice vectors}
\STATE \textbf{return} $P$
\end{algorithmic}
\end{algorithm}

\paragraph{Complexity estimates.} For completeness, let us also give a high-level description of the main argumentation for the time and space complexities of lattice sieving. First, if $\vc{v}, \vc{w}$ have equal Euclidean norms (similar norms occur often in high-dimensional sieving instances), then the difference vector $\vc{u} = \vc{v} - \vc{w}$ is a shorter vector than $\vc{v}, \vc{w}$ if and only if $\vc{v}, \vc{w}$ have a mutual angle $\phi < \frac{\pi}{3}$. The probability of this occurring, for e.g.\ uniformly random vectors $\vc{v}, \vc{w}$ of unit length, is proportional to $\sin(\frac{\pi}{3})^d = (\frac{3}{4})^{d/2}$ due to volume arguments of hyperspherical caps. To guarantee that the next generation has approximately the same size as the previous one, so that after a potentially large number of iterations we will not end up with an empty list, we need $|P'| \approx |P|^2 \cdot (\frac{3}{4})^{d/2} \approx |P|$, as there are $|P|^2$ pairs of parents in $P$, and they produce good offspring (a short difference vector) with probability approximately $(\frac{3}{4})^{d/2}$. Solving for $|P|$ gives $|P| \approx (\frac{4}{3})^{d/2} \approx 2^{0.208d}$ as an estimate of the memory complexity (population size) needed to succeed, and since all pairs of vectors are considered for generating offspring, this gives a quadratic time complexity scaling as $|P|^2 \approx (\frac{4}{3})^{d} \approx 2^{0.415d}$, as argued in e.g.\ ~\cite{nguyen08}.\footnote{The number of applications of the sieve necessary to converge to a solution, is only polynomial in $d$, and negligible compared to the exponential running time for each application of the sieve.}


\section{PAST SIEVING TECHNIQUES}
\label{sec:tech1}

The previous section described how the most basic lattice sieving algorithm could be viewed as an evolutionary algorithm. Over time, various improvements have been proposed for lattice sieving, and many of them naturally relate to techniques that have been previously studied in the context of evolutionary computation as well.

\paragraph{Multi-parent offspring and tuple lattice sieving.} A technique discussed in e.g.~\cite{eiben94} is the relevance of scenarios where offspring is produced by more than two parents. By selecting more parents and inheriting the best genes from all of them, stronger offspring can sometimes be generated than with two parents. This idea has been studied in the context of lattice sieving as \textit{tuple lattice sieving}~\cite{bai16,herold18}, where tuples of up to $k \geq 2$ vectors are recombined to generate shorter lattice vectors. In general, this approach leads to better memory complexities than the standard approach (smaller populations suffice to guarantee a productive evolution process) but to worse time complexities till convergence (finding suitable $k$-tuples of parents for recombination requires more work).

\paragraph{Genetic segregation and nearest neighbor searching.} A common technique in evolutionary algorithms, related to \textit{niching} and \textit{speciation}, is to subdivide the search space into mutually disjoint regions, and letting different subspecies of the population coevolve separately, with recombinations happening only within each subpopulation. This closely relates to the application of techniques from \textit{nearest neighbor searching}~\cite{indyk98} to lattice sieving~\cite{laarhoven15crypto,becker16lsf,laarhoven16phd}. By dividing the high-dimensional search space into regions, and separately recombining parents in each of these region, one generally still finds good parents for producing better offspring, while saving a lot of time on attempting to mate parents which are unsuitable couples for reproduction.

\paragraph{Progressive preferences and progressive lattice sieving.} For finding optimal solutions in the entire search space, some methods in EA have proposed a progressive approach, where initially only a subset of the constraints (search space) is studied to find local solutions, before expanding to a wider search space and finding global solutions~\cite{coello07}. Similar ideas have recently been explored in the context of lattice sieving~\cite{laarhoven18pqcrypto,ducas18}, starting to sieve in a sublattice of the original lattice before widening the search space. This heuristically and practically accelerates the time until convergence.

\paragraph{Island models and parallelization.} When attempting to parallelize evolutionary algorithms, one naturally faces the question how to subdivide the population for separate processing, and how to then merge the local results to find global optima~\cite{back00b}. For lattice sieving, this topic has been studied in e.g.\ \cite{mariano17,albrecht19}, where initially parts of the population were recombined on individual nodes, and the individuals then migrated across different nodes to guarantee reproduction with all suitable mates.

\paragraph{Crowding and replacing parents directly.} Perhaps the most common method in evolutionary computation for selecting the next generation is to generate offspring, sort all individuals by their fitness, and then take the fittest ones for the next generation. Lattice sieving commonly applies a form of \textit{crowding}, where each time a child is produced, one of its parents is discarded for the next generation. This guarantees that each `corner' of space only contains few vectors, and allows for the complexity estimates from the previous section based on pairwise angles between vectors. 

\section{NEW SIEVING TECHNIQUES}
\label{sec:tech2}

Besides existing methods from EA, which have already been studied in the context of sieving, perhaps of more interest are those techniques that have not yet been considered to improve sieving algorithms. Below we consider three main concepts, which will then be evaluated experimentally as well.

\paragraph{Encodings.} Commonly, evolutionary algorithms do not work directly with the population, but with \textit{encodings} of the population that allow for more natural, genetic recombinations and mutations. Besides the direct application of evolutionary techniques to lattice sieving, working with lattice vectors in terms of their coordinates, one could also encode members of the population differently, i.e.\ by encoding a lattice vector $\vc{v}$ by its coefficient vector $\vc{\lambda}$ in the basis $\vc{B}$: 
\begin{align*}
    \vc{v} = (v_1, \dots, v_d) = \sum_{i=1}^d \lambda_i \vc{b}_i \quad \stackrel{\text{encode}}{\longrightarrow} \quad \vc{\lambda} = (\lambda_1, \dots, \lambda_d).
\end{align*}
Recombining vectors can be done the same as before, where subtracting two coordinate vectors equivalently corresponds to subtracting their coefficient vectors in terms of the basis. In some cases it may be more convenient to work with the vectors directly (i.e.\ to compute fitness), but for mutations discussed below, working with encodings is more natural.

\paragraph{Mutations.} In existing lattice sieving methods, population members never undergo any unitary mutations. Note that the common technique of modifying single genes of an individual does not quite make sense for lattice sieving: by modifying a coordinate, one may step outside the lattice, and the algorithm will no longer solve the right problem. With the encoding described above, mutations can be applied more naturally by sometimes adding random, small noise to generated offspring, which would correspond to adding short combinations of basis vectors. Note that as lattice sieving algorithms progress, such short combinations of basis vectors are typically much longer than the generated children, and mutations generally decrease the fitness; if at all useful, mutations would have to be done very sporadically.

\paragraph{Global selection.} Finally, existing lattice sieving approaches always do updates to the population on a local scale: one or both of the parents are replaced by the children, to guarantee that the list has the pairwise reduction property outlined in Section~\ref{sec:evol} for arguing heuristic bounds on the population size. Instead, it may be beneficial to keep parents more often, and to make a final selection for the next generation on a global scale: order all generated offspring and parents by their fitness, and keep the strongest members.


\section{EXPERIMENTS}
\label{sec:exp}

\begin{figure}[!t]
    \centering
    \subfloat[][Fitness levels (average/minimum) per generation]
        {\includegraphics[width=7.3cm]{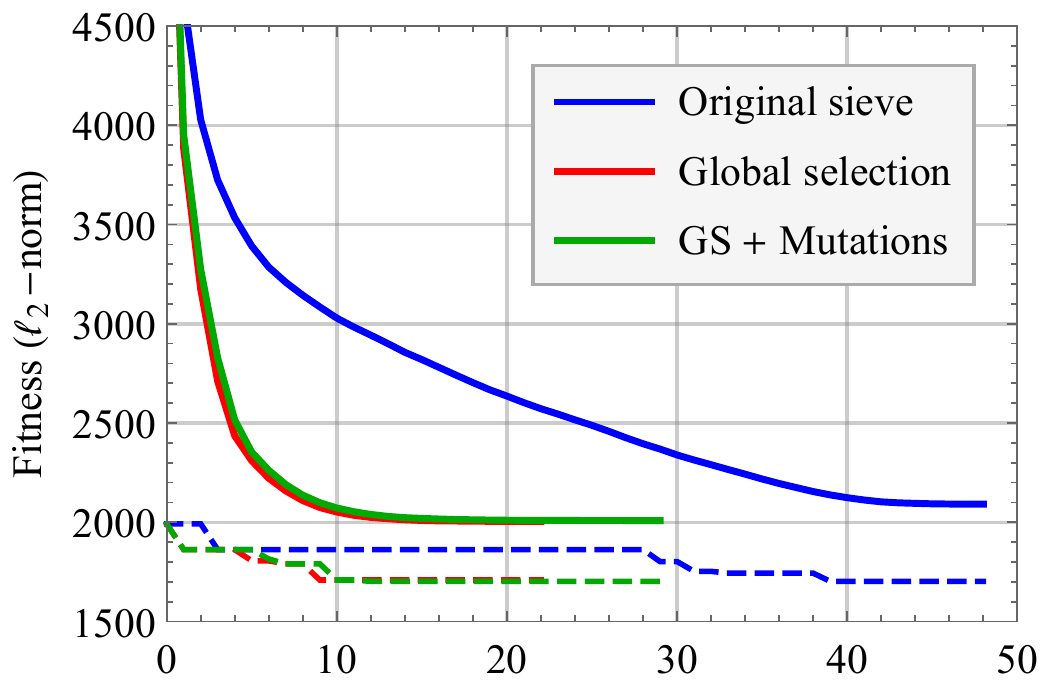}
    \label{fig:exp1}} \\
    \subfloat[][Replaced vectors (cumulative/separated) per generation]
        {\includegraphics[width=7.3cm]{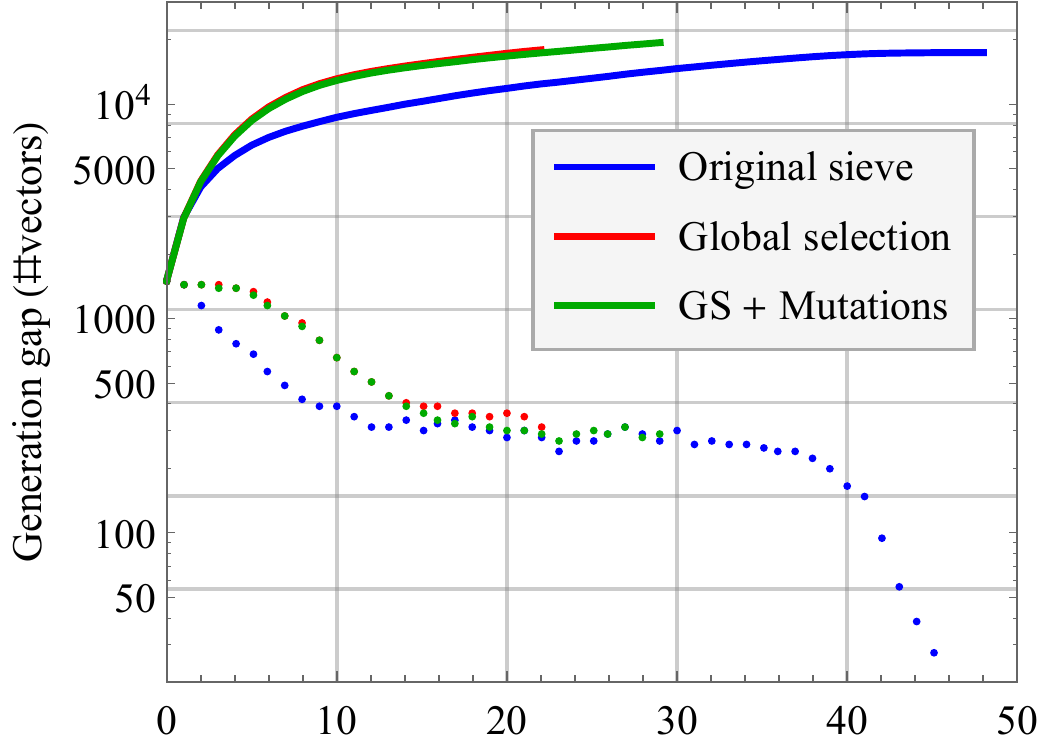}
    \label{fig:exp2}} \\
    \caption{\textbf{(Figure \ref{fig:exp1})} depicts the average fitness (\textbf{thick}) and best fitness (dashed) per generation -- lower $\ell_2$-norms correspond to a higher fitness level. \textbf{(Figure \ref{fig:exp2})} depicts the number of surviving children per generation (generation gap) -- the higher the gap, the faster the evolution. Both graphs depict the original sieve ({\color{blue}blue}), the sieve with global updates ({\color{red}red}), and the sieve with both global selection and occasional mutations ({\color{green!50!black}green}).}
    \label{fig:exp}
\end{figure}

We have implemented the basic lattice sieving approach outlined in Section~\ref{sec:evol}, as well as (1) the concept of global selection for lattice sieving (rather than local replacements), and (2) mutations based on the representation of vectors in the lattice basis. We have tested these algorithms on a $40$-dimensional random lattice from the SVP challenge website~\cite{svp}, starting with an LLL-reduced basis of the lattice and letting the algorithm run for a number of generations, until no more progress is made between successive generations. The initial population size was set to $1500$ lattice vectors, and each successive generation was also limited to using a population of at most this size. Figure~\ref{fig:exp1} shows the experimental results of the average (and minimum) fitness per generation against the generation number, and below we discuss the effects of these modifications in more detail.

\paragraph{Original approach.} For the original sieving approach, we look for pairs $\vc{v}, \vc{w} \in P$ such that $\vc{u} = \vc{v} - \vc{w}$ is a shorter vector than one of its parents. If this is the case, we replace the longer of the two parent vectors with this child, and we continue until all (unmodified) parent vectors have been compared to see if their offspring leads to an improvement in the population. This is the basic algorithm from~\cite{nguyen08,laarhoven16phd}, and the blue line in Figure~\ref{fig:exp1} shows how the average norms of the vectors in the population decrease over time. The dashed blue line further shows the progression in terms of the norm of the shortest member of the population, and in our example run it took $39$ generations for the algorithm to find the shortest non-zero vector $\vc{s} \in \cL$ of norm $\|\vc{s}\| \approx 1702$. After $48$ generations and $20$ seconds the algorithm terminated with an average fitness ($\ell_2$-norm) of $2092$, and a final population size of $1386$ lattice vectors, having done $17300$ updates to the population.

\paragraph{Global selection.} For this variant, we considered all children $\vc{u} = \vc{v} - \vc{w}$ for potential insertion in the population for the next generation, and at the end we simply selected the $1500$ shortest/fittest members for survival. As the red curves in Figure~\ref{fig:exp} show, this decreases the number of generations needed for convergence, the average norm of vectors decreases faster between generations, and the vectors get replaced more quickly than before. After $22$ generations and $1.5$ seconds on the same machine, the algorithm converged to a population of $1500$ short lattice vectors of average length $2004$, having done $17800$ vector replacements. Unfortunately, the algorithm failed to find a shortest vector, and the shortest vector in the final population had norm $1709$.

\paragraph{Mutations.} For this variant, not depicted in Figure~\ref{fig:exp}, we performed sieving with local updates (children replacing their direct parents), but we added occasional mutations of children -- in $10\%$ of all cases we added/removed single basis vectors to children to create more diversity in the population. 

However, in combination with local updates, mutations are not very successful, and over time the average norm of vectors in the population only increased. This is inherent to the local updates, where mutations commonly increase the norm and lead to a local update to the population that only leads to longer lattice vectors. With a global survivor selection procedure, bad mutations can be filtered out for the next generation, but with local updates this is not the case.

\paragraph{Global selection and mutations.} The final variant in our experiments uses both global selection and occasional mutations of children, described above. In this particular example, adding mutations to the global selection sieve indeed helped: after $29$ generations and $1.9$ seconds, we converged to a population of $1500$ lattice vectors with average Euclidean norm $2008$ and minimum norm $1702$, corresponding to the shortest non-zero vector in the lattice (having done $19300$ updates to the population). This in contrast with the earlier global selection algorithm without mutations, where we converged to a local solution of larger norm $1709$. The time till convergence as well as the number of generations needed for convergence are a bit worse compared to not using mutations, but we did find the optimal solution with this variant.


\paragraph{Discussion.} The results in Figure~\ref{fig:exp1} demonstrate what we might expect to happen when using these modifications. Using a global survivor selection approach (rather than the local replacements in existing sieving algorithms), the overall quality of the population improves faster in each generation, and the algorithm converges more quickly towards an optimal solution (i.e.\ a shortest non-zero vector of the lattice). With only the global selection modification we further noticed we were ``unlucky'' in converging to a local solution of norm $1709$, rather than the shortest vector in the lattice of norm $1702$. With the extra randomness generated by the genetic mutations, we did eventually find the shortest vector in our population, although the number of iterations till convergence increased slightly. We expect this behavior to appear in other examples too: mutations commonly will not increase the performance, but may help in preventing convergence towards local optima.

\paragraph{On the absence of global selection in sieving.} As the idea of population-wide survivor selection appears very natural, and various more advanced techniques have already been considered in the context of lattice sieving, one might wonder why this idea has not yet been applied to sieving. Perhaps the main reason for this is that the complexity estimates of lattice sieving, described in Section~\ref{sec:evol}, crucially rely on the population having the property that any two vectors $\vc{v}, \vc{w} \in P$ have a pairwise angle of at least $\frac{\pi}{3}$; otherwise, we would find the child $\vc{u} = \vc{v} - \vc{w}$ as a shorter child, and one of the parents would have been replaced with $\vc{u}$. Given that all pairs of vectors are relatively far apart in terms of their pairwise angles, this then allows us to use sphere packing bounds~\cite{nguyen08} to obtain heuristic upper bounds on the population size and, consequently, on the running time of the algorithm. When we do updates globally, and select only the fittest members for the next generation, we no longer have these heuristic guarantees for the time and space complexities of sieving. So even though in practice, as our experiments indicated, this global selection modification only appears to improve the population quality, from a complexity-theoretic point of view this modification is somewhat counter-intuitive.

\section{CONCLUSION}
\label{sec:disc}

In this paper we demonstrated a new, natural connection between lattice sieving algorithms used in cryptanalysis on the one hand, and techniques in evolutionary algorithms on the other hand. We analyzed how ideas and terminology in both fields relate, and how certain ideas from EA that have not yet been applied to lattice sieving may be of interest for improving sieving algorithms. In particular, the idea of a global selection procedure appears promising, and although from a certain point of view this modification is somewhat unnatural, experiments suggest that this may well benefit the performance of lattice sieving in practice. Note that we have only tested these modifications with a basic sieve in a low-dimensional lattice ($d = 40$), and analyzing how this modification interacts with other existing improvements and tweaks to state-of-the-art lattice sieving implementations (see e.g.\ \cite{albrecht19}) is left for future work.

Part of the aim of this work is also to stimulate a further exchange of ideas between both fields, as several existing ideas which have turned out to be useful in lattice sieving have been studied in the context of evolutionary computation long ago, and may well have been introduced to lattice sieving sooner, had ideas between both fields been exchanged sooner. Interested readers from the area of AI may wish to refer to \cite{laarhoven16phd} for an overview of lattice sieving techniques; to \cite{becker16lsf} for the current theoretical state-of-the-art in terms of lattice sieving; and to \cite{albrecht19} for what is currently (as of early 2019) the fastest lattice sieving method in practice. Given the similarities between lattice sieving and evolutionary computation, there may well be further ways to improve lattice sieving with existing techniques from AI. 

Besides the relation with lattice sieving discussed here, some other techniques in the broader field of cryptanalysis also follow a similar procedure of (1) generating a random, large population; (2) combining members in this population to form better solutions; and (3) ultimately finding a solution in the final population. We explicitly state two examples:
\begin{itemize}
    \item \textbf{The Blum--Kalai--Wasserman (BKW) algorithm.} \\
    One of the fastest known methods for attacking cryptographic schemes based on the hardness of learning parity with noise (LPN) and learning with errors (LWE)~\cite{regev05,regev06} is the BKW algorithm~\cite{blum03}. From a high-level point of view, one starts with a list of integer vectors, and tries to find short combinations that cancel out many of the coordinates, thus leading to vectors with many zeros.
    \item \textbf{Decoding random (binary) linear codes.} \\ 
    For understanding the security of state-of-the-art code-based cryptographic schemes~\cite{mceliece78,bernstein09}, the fastest known attacks solve a decoding problem for random binary, linear codes. These also commonly start by generating a large population of $\{0,1\}$-strings, and then forming combinations to cancel out many of the coordinates and obtain a vector with low Hamming weight~\cite{may15}. 
\end{itemize}
Both approaches can similarly be interpreted as evolutionary algorithms, and we leave a further study of this relation for future work.

\section*{ACKNOWLEDGMENTS}
The author is supported by a Veni Innovational Research Grant from NWO under project number 016.Veni.192.005.

\end{document}